\PassOptionsToPackage{unicode}{hyperref}
\PassOptionsToPackage{hyphens}{url}
\PassOptionsToPackage{dvipsnames,svgnames,x11names}{xcolor}
\documentclass[
]{article}
\usepackage{amsmath,amssymb}
\usepackage{lmodern}
\usepackage{iftex}
\ifPDFTeX
  \usepackage[T1]{fontenc}
  \usepackage[utf8]{inputenc}
  \usepackage{textcomp} 
\else 
  \usepackage{unicode-math}
  \defaultfontfeatures{Scale=MatchLowercase}
  \defaultfontfeatures[\rmfamily]{Ligatures=TeX,Scale=1}
\fi
\IfFileExists{upquote.sty}{\usepackage{upquote}}{}
\IfFileExists{microtype.sty}{
  \usepackage[]{microtype}
  \UseMicrotypeSet[protrusion]{basicmath} 
}{}
\makeatletter
\@ifundefined{KOMAClassName}{
  \IfFileExists{parskip.sty}{%
    \usepackage{parskip}
  }{
    \setlength{\parindent}{0pt}
    \setlength{\parskip}{6pt plus 2pt minus 1pt}}
}{
  \KOMAoptions{parskip=half}}
\makeatother
\usepackage{xcolor}
\usepackage{graphicx}
\makeatletter
\def\maxwidth{\ifdim\Gin@nat@width>\linewidth\linewidth\else\Gin@nat@width\fi}
\def\maxheight{\ifdim\Gin@nat@height>\textheight\textheight\else\Gin@nat@height\fi}
\makeatother
\setkeys{Gin}{width=\maxwidth,height=\maxheight,keepaspectratio}
\makeatletter
\def\fps@figure{htbp}
\makeatother
\NewDocumentCommand\citeproctext{}{}
\NewDocumentCommand\citeproc{mm}{%
  \begingroup\def\citeproctext{#2}\cite{#1}\endgroup}
\makeatletter
 \let\@cite@ofmt\@firstofone
 \def\@biblabel#1{}
 \def\@cite#1#2{{#1\if@tempswa , #2\fi}}
\makeatother
\newlength{\cslhangindent}
\newlength{\csllabelwidth}
\newenvironment{CSLReferences}[2] 
 {\begin{list}{}{%
  \setlength{\itemindent}{0pt}
  \setlength{\leftmargin}{0pt}
  \setlength{\parsep}{0pt}
  \ifodd #1
   \setlength{\leftmargin}{\cslhangindent}
   \setlength{\itemindent}{-1\cslhangindent}
  \fi
  \setlength{\itemsep}{#2\baselineskip}}}
 {\end{list}}
\usepackage{calc}

\ifLuaTeX
\usepackage[bidi=basic]{babel}
\else
\usepackage[bidi=default]{babel}
\fi
\babelprovide[main,import]{american}

\def\languageshorthands#1{}
\ifLuaTeX
  \usepackage{selnolig}  
\fi
\IfFileExists{bookmark.sty}{\usepackage{bookmark}}{\usepackage{hyperref}}
\IfFileExists{xurl.sty}{\usepackage{xurl}}{} 
\hypersetup{
  pdftitle={madupite: A High-Performance Distributed Solver for
Large-Scale Markov Decision Processes},
  pdfauthor={Matilde Gargiani, Philip Pawlowsky, Robin Sieber, Václav
Hapla, John Lygeros},
  pdflang={en-US},
  colorlinks=true,
  linkcolor={Maroon},
  filecolor={Maroon},
  citecolor={Blue},
  urlcolor={Blue},
  pdfcreator={LaTeX via pandoc}}

\title{madupite: A High-Performance Distributed Solver for Large-Scale
Markov Decision Processes}

\definecolor{c53baa1}{RGB}{83,186,161}
\definecolor{c202826}{RGB}{32,40,38}

\usepackage[affil-it]{authblk}
\usepackage{orcidlink}
\author[1%
  *%
  \ensuremath\mathparagraph]{Matilde Gargiani%
    \,\orcidlink{0000-0001-8615-6214}\,%
    }
\author[1%
  *%
  \ensuremath\mathparagraph]{Philip Pawlowsky%
    \,\orcidlink{0009-0003-9732-3884}\,%
    }
\author[1%
  *%
  \ensuremath\mathparagraph]{Robin Sieber%
    \,\orcidlink{0009-0002-8592-8387}\,%
    }
\author[2,3%
  ]{Václav Hapla%
    \,\orcidlink{0000-0002-9190-2207}\,%
    }
\author[1%
  \ensuremath\mathparagraph]{John Lygeros%
    \,\orcidlink{0000-0002-6159-1962}\,%
    }

\affil[1]{Automatic Control Laboratory (IfA), ETH Zurich, 8092 Zurich,
Switzerland%
  }
\affil[2]{Department of Earth and Planetary Sciences, ETH Zurich, 8092
Zurich, Switzerland%
  }
\affil[3]{Department of Applied Mathematics, FEECS at VSB-TU Ostrava,
Czechia%
  }
\affil[$\mathparagraph$]{Corresponding author: %
}
\affil[*]{These authors contributed equally.}
\date{9 September 2024}

\begin{document}
\maketitle

\begin{figure}
\centering
\includegraphics[width=0.4\textwidth,height=\textheight]{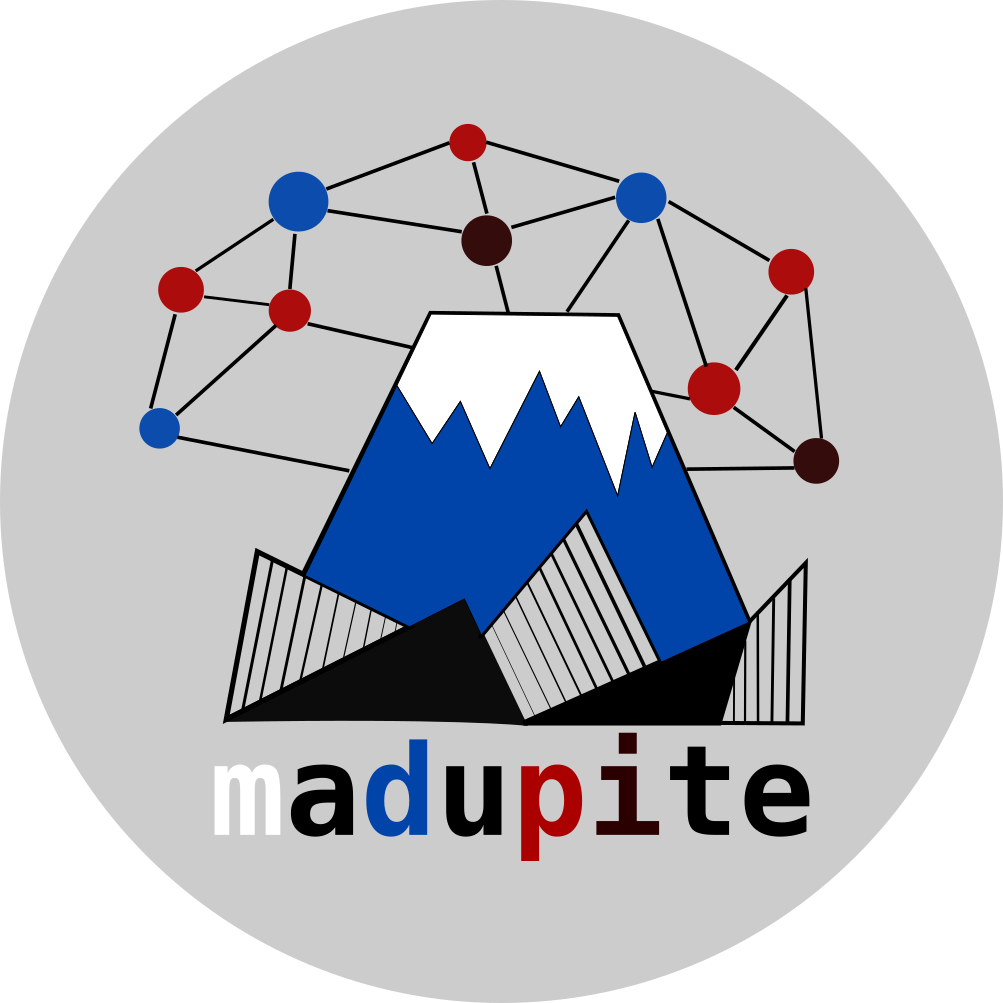}
\caption{Logo of \texttt{madupite}.}
\end{figure}

\section{Summary}\label{summary}

We propose \texttt{madupite}, a distributed high-performance solver for
Markov Decision Processes (MDPs). MDPs are a powerful mathematical tool
to model a variety of problems arising in different fields
(\citeproc{ref-mdp_methods_applications}{Feinberg \& Shwartz, 2002};
\citeproc{ref-realworld_mdp}{White, 1985}), from finance
(\citeproc{ref-mdp_finance}{Bäuerle \& Rieder, 2011}) to epidemiology
(\citeproc{ref-mdp_epidemiology}{Steimle \& Denton, 2017}) and traffic
control (\citeproc{ref-mdp_trafficcontrol}{Xu et al., 2016}). In general
terms, MDPs are used to mathematically characterize dynamical systems
whose state is evolving in time as a consequence of actions that we play
on the system and disturbances that are acting on it. The goal is
generally to select actions in order to minimize in expectation a
certain cumulative discounted metric over time, \emph{e.g.}, deviations
from a reference state and/or costs incurred by selecting certain
actions in specific states of the system
(\citeproc{ref-bellman_book}{Bellman, 1957};
\citeproc{ref-bertsekas_book_2}{Bertsekas, 2007}).

MDPs arising from real-world applications tend to be extremely
high-dimensional and in some cases the number of states in the system
grows exponentially with the number of certain parameters
(\citeproc{ref-powell_approximate_2011}{Powell, 2011}). This phenomenon
is known as \emph{curse-of-dimensionality} and it is generally tackled
in reinforcement learning with the deployment of function approximations
(\citeproc{ref-sutton_RL}{Sutton \& Barto, 2018}). The deployment of the
latter leads to a smaller size optimization problem since, instead of
optimizing for the number of states, there is only need to optimize for
the number of parameters deployed in the function approximation, which
is generally much smaller than the original state space size. This comes
at the price of introducing sub-optimality with respect to the original
solution.

\section{Statement of need}\label{statement-of-need}

Modern high-performance clusters and super-computers offer the
possibility of simulating, storing and solving large-scale MDPs. To
exploit modern computational resources, high-performance software
packages which can efficiently distribute computation as well as scale
adequately are needed. Even though there are a number of toolboxes to
solve MDPs, such as \texttt{pymdptoolbox}
(\citeproc{ref-mdptoolbox}{Chadès et al., 2014}) and the recent
\texttt{mdpsolver} (\citeproc{ref-mdp_solver}{Reenberg Andersen \& Fink
Andersen, 2024}), to the best of our knowledge there is no existing
solver that combines high-performance distributed computing with the
possibility of selecting a solution method that is tailored to the
application at hand. \texttt{pymdptoolbox} is coded in plain Python,
which is limited in scalability due to the lack of parallel support. In
addition, it does not support parallel and distributed computing.
\texttt{mdpsolver} is instead written in C++, supports parallel
computing and comes with a user-friendly Python API. Even so,
\texttt{mdpsolver} is not fully featured. The solution methods available
are limited to modified policy iteration, a dynamic programming (DP)
method which has shown poor performance for a significant class of
problems of practical interest
(\citeproc{ref-gargiani_igmrespi_2023}{Gargiani et al., 2023},
\citeproc{ref-gargiani_ipi_2024}{2024}). In addition, \texttt{mdpsolver}
makes certain implementation choices that limit its applicability;
\emph{e.g.}, matrices with values and indices being stored in nested
\texttt{std::vector} independently of their sparsity degree and thus
precluding the use of available optimized linear algebra routines.

Our solver aims to enable the solution of large-scale MDPs with more
than a million states through efficient utilization of modern
computational resources, while abstracting away the complexity of
distributed computing to provide a user-friendly experience.
\texttt{madupite} is a high-performance distributed solver which is
capable of efficiently distributing the memory load and computation,
comes with a wide range of choices for solution methods enabling the
user to select the one that is best tailored to its specific
application, and, last but not least, its core is in C++ but it is
equipped with a user-friendly Python API to enable its deployment to a
broader range of users. The combination of distributed computing,
user-friendly Python API, and support for a wide range of methods in
\texttt{madupite} will enable researchers and engineers to solve exactly
gigantic scale MDPs which previously could only be tackled via function
approximations.

\section{Problem Setting and Solution
Methods}\label{problem-setting-and-solution-methods}

\texttt{madupite} solves infinite-horizon discounted MDPs with finite
state and action spaces. This problem class can be efficiently tackled
with \emph{inexact policy iteration methods} (iPI)
(\citeproc{ref-gargiani_igmrespi_2023}{Gargiani et al., 2023},
\citeproc{ref-gargiani_ipi_2024}{2024}). These methods are a variant of
policy iteration (\citeproc{ref-bertsekas_book_2}{Bertsekas, 2007}),
where inexactness is introduced at the policy evaluation step for
scalability reasons. iPI methods are general enough to embrace standard
DP methods, such as value iteration and modified policy iteration, which
are the main solution methods of \texttt{mdpsolver}. The interested
readers should refer to (\citeproc{ref-gargiani_ipi_2024}{Gargiani et
al., 2024}) for an in-depth description of the problem setting and a
thorough mathematical analysis of iPI methods. The versatility of iPI
methods makes them particularly suited to efficiently solve different
large-scale instances, while their structure is also favorable for
distributed implementations, which are needed to exploit
high-performance computing clusters.

The great flexibility of \texttt{madupite} is that on the algorithmic
side it relies on the possibility of customizing the iPI method deployed
for the specific problem at hand by tuning the level of inexactness and
selecting among a wide range of inner solvers for the approximate policy
evaluation step (step 8 of Algorithm 3 in
(\citeproc{ref-gargiani_ipi_2024}{Gargiani et al., 2024})). It is indeed
empirically and theoretically demonstrated that, depending on the
specific structure of the problem, different inner solvers may enhance
the convergence performance
(\citeproc{ref-gargiani_igmrespi_2023}{Gargiani et al., 2023},
\citeproc{ref-gargiani_ipi_2024}{2024}).

\section{Implementation}\label{implementation}

The core of \texttt{madupite} is written in C++ and relies on PETSc
(Portable, Extensible Toolkit for Scientific Computation) for the
distributed implementation of iPI methods
(\citeproc{ref-petsc-efficient}{Balay et al., 1997},
\citeproc{ref-petsc-web-page}{2024a},
\citeproc{ref-petsc-user-ref}{2024b}). PETSc is an open-source
high-performance C library and comes with a wide range of
highly-optimized linear system solvers and memory-efficient sparse
linear algebra data types and routines that can be used for a variety of
problems, including DP. We rely on PETSc also as it natively adopts a
distributed memory parallelism using the MPI standard to enable
scalability beyond one CPU. MPI allows users to abstract the parallelism
away from the underlying hardware and enables them to run the same code
in parallel on their personal device or even on multiple nodes of a
high-performance computing cluster. \texttt{madupite} itself is a
fully-featured C++20 library and, by leveraging the \texttt{nanobind}
binding library (\citeproc{ref-nanobind}{Jakob, 2022}), offers an
equivalent API in Python that can be installed as a package using
\texttt{pip}.

Finally, \texttt{madupite} allows the user to create an MDP by loading
offline data collected from previously run experiments as well as from
online simulations, offering the possibility of carrying out both the
simulations and the solution in a completely parallel/distributed
fashion. More details on how to use \texttt{madupite} and all of its
functionalities can be found in its documentation and in the
\texttt{examples} folder, where we provide an extensive selection of
code-examples on how to use this library in Python and C++.

\section{Acknowledgements}\label{acknowledgements}

This work was supported by the European Research Council under the
Horizon 2020 Advanced under Grant 787845 (OCAL) and by the SNSF through
NCCR Automation (Grant Number 180545).

\section*{References}\label{references}
\addcontentsline{toc}{section}{References}

\phantomsection\label{refs}

\end{document}